# 2D superconductivity and vortex dynamics in 1T-MoS$_2$

Chithra H. Sharma[1], Ananthu P. Surendran[1], Sangeeth S. Varma[2] and Madhu Thalakulam[1*]

[1]School of Physics, Indian Institute of Science Education & Research Thiruvananthapuram, Kerala, 695551 India

[2] BITS-Pilani , K. K. Birla Goa Campus, Goa, 403726 India

Two-dimensional (2D) superconductivity is a fascinating phenomenon packed with rich physics and wide technological application. The vortices and their dynamics arising from classical and quantum fluctuations give rise to Berezinskii-Kosterlitz-Thouless (BKT) transition and 2D Bose metallic phase both of which are of fundamental interest. In 2D, observation of superconductivity and the associated phenomena are sensitive to material disorders. Highly crystalline and inherently 2D van der Waals (vW) systems with carrier concentration and conductivity approaching metallic regime have been a potential platform. The metallic 1T phase of MoS$_2$, a widely explored vW material system controllably, engineered from the semiconducting 2H phase, is a tangible choice. Here, we report the observation of 2D superconductivity accompanied by BKT transition and Bose metallic state in a few-layer 1T-MoS$_2$. Structural characterization shows excellent crystallinity over extended lateral dimension. The electrical characterization confirms the metallic nature down to 4 K and a transition to a superconducting state below 1.2 K with a T$_c$ ~ 920 mK. The 2D nature of the superconducting state is confirmed from the magneto-transport anisotropy against field orientations and the presence of BKT transition. In addition, our sample showcases a manifold increase in the parallel upper-critical-field above the Pauli limit. The inherent two-dimensionality and possibility of scalably engineering semiconducting, metallic and superconducting phases makes MoS$_2$ a potential candidate for hosting monolithic all-two-dimensional hybrid quantum devices.

## Introduction

Systems with reduced dimensions provide an opportunity to test fundamental theorems, formulate new ones and access new technologies. The search for non-Abelian quantum Hall systems in one-dimensional constrictions[1] and superconductivity in two-dimensions (2D)[2,3] are a few to mention. Mermin-Wagner theorem forbids any kind of ordering and spontaneous symmetry breaking in dimensions $D \leq 2$ for non-zero temperatures[4]. Unlike in three-dimensions, disorder precludes any long-range order in reduced dimensions. In contrast, experiments involving 2D systems vary; isolation of mono-layer of graphite[5], 2D metallic state at ultra-low-temperatures[6] and 2D superconducting phase transitions[3] invoke discussions on the limit and applicability of the theorem to practical systems. Traditionally, 2D systems are realized on interfaces of bulk materials or thin films which are strongly disordered while, controlled experiments to make fundamental investigations in this direction require systems with minimal structural disorder. van der Waals (vW) materials, highly crystalline naturally confined 2D systems in the few-layer limit, have recently become the ubiquitous choice for exploring various quantum phenomena[7-11]. Among the vW materials, transitional metal dichalcogenides (TMDCs), especially MoS$_2$, have shown potential as a platform for hosting mesoscopic quantum devices such as quantum point contacts and quantum dots[8,9,12]. Coherent electrical transport, quantum oscillations, charge and conductance quantization, weak localization etc., have been observed in $MoS_2$[7-11]. Apart from the high degree of electrical tunability, what makes $MoS_2$ fascinating is the presence of 2D polymorphs displaying a wide spectrum of electrical properties[13]; 2H and 3R are large bandgap semiconductors, 1T' and 1T" are narrow bandgap semiconductors and, 1T is metallic. Prospect of a 2D superconducting polymorphic phase and the possibility of engineering it controllably and scalably could make $MoS_2$ a universal platform for hosting hybrid quantum circuits consisting of semiconducting, metallic and superconducting regions; 2D Josephson junction circuits[14] and cavity QED circuits consisting semiconducting and superconducting circuit elements[15] are a few to mention. Recently superconducting phase transition has been reported in 1T and 1T' $MoS_2$[16-18], though, the electrical transport is not explored in detail and the dimensionality of superconductivity is not addressed. 2D superconductivity is limited to systems with sheet resistance less than the resistance quantum, $R_Q = \frac{h}{4e^2} \sim 6.45 \text{ k}\Omega$[3,19]; metallic nature and high carrier concentration make 1T-$MoS_2$ a natural choice.

* madhu@iisertvm.ac.in

Superconductivity in the 2D limit is featured by the presence of vortices and their dynamics. Berezinskii-Kosterlitz-Thouless (BKT) phase transition[20,21] caused by vortex-antivortex unbinding and, the anisotropy in the magneto-transport with respect to the magnetic field (B-field) orientations are considered as the touchstone features of 2D superconductivity. In systems with low disorder, a 2D quantum metallic phase, the Bose metal, is emerged due to magnetic field induced gauge fluctuations[22-24], as opposed to an insulating state observed in strongly disordered systems. Until the advent of vW materials[3] 2D superconductivity has been confined to disordered material systems[19,25-28] such as interfaces of bulk materials and thin films. Low structural disorders and high crystallinity makes 2D superconductivity and associated phenomena easier to observe in vW systems compared to disordered 2D systems. A substantial increase in the parallel upper critical-field compared to that predicted by the Pauli limit has been observed in crystalline 2D systems[25,29-31]. This has been attributed to spin-orbit coupling (SOC) and broken-inversion symmetry resulting spin-splitting of conduction electrons making it resilient to B-field induced pair breaking. A similar enhancement in the upper critical field also has been reported in centro-symmetric systems recently[32].

Owing to the limited electrical tunability, attempts to explore superconductivity in Graphene by doping[33] have not gained much inroads. Many of the transition metal carbides and nitrides exhibit superconductivity in the bulk form[34] and 2D superconductivity has been identified in $Mo_2C$ recently[35]. Among the TMDCs, $NbSe_2$ and $TaS_2$ have shown superconductivity in the bulk[36,37] and in the 2D limit[30,38]. Both mono- and few-layer $NbSe_2$ have shown 2D superconductivity with Ising pairing[30] and evolution of Bose metallic phase[22]. $NbSe_2$ and $TaS_2$ are metallic, with limited electrical tunability and prone to oxidation, making device fabrication and integration with other materials challenging[22,38,39]. Conventional gating technique is inadequate to achieve the required carrier concentration to observe 2D superconductivity and, ionic liquid gating technique is employed instead. By this method, carrier densities $\sim 10^{14}$ cm$^{-2}$ leading to a transition from the insulating to the 2D superconducting state has been reported in 2H-$MoS_2$[29,40-42], 2H-$WS_2$[31,43] and 1T-$SnSe_2$[44]. In addition, Ising pairing has been observed in ionic-liquid gated 2H-$MoS_2$[29,41,42] and 2H-$WS_2$[31,43] devices. Unlike the traditional dielectric gating, ionic liquid gating technique has limited scalability and may fall short of the repeatability and robustness required for device technology.

1T-$MoS_2$, metallic in nature with high carrier concentration, qualifies the resistance requirement to exhibit 2D superconductivity. 1T phase is generally prepared by structural modification of the 2H phase either by chemical intercalation[45] or by physical means using high-energy electrons or plasma treatment[46-48]. Stability of the 1T phase is a highly debated topic. Chemically converted 1T $MoS_2$ is reported to be unstable against time and temperature making it not suitable for the device fabrication, in contrast to those prepared by physical routes[47-50]. Scalable and controllable engineering of 1T phase on the 2H phase has been demonstrated recently[48]. The possibility of observing 2D superconductivity on the 1T phase can make $MoS_2$ a potential choice for hosting 2D monolithic hybrid quantum circuits.

In this manuscript, we report the observation of 2D superconductivity featuring BKT transition and evolution of Bose metallic state in an ~ 8nm thick (~12 layers) 1T-$MoS_2$ sample. The 1T-$MoS_2$ samples are prepared from the 2H phase using forming-gas microwave plasma assisted phase engineering. We characterize the 1T phase by high-resolution transmission electron microscopy (HR-TEM). Details of the sample preparation and in-depth structural characterization can be found elsewhere[48]. The electrical characterizations consist of low-noise four-probe (4P) transport measurements from 300 K down to 4 K and, magneto-transport measurements from 4 K down to 12 mK. Current-voltage (I-V) characteristics show excellent linearity down to 4 K void-of any dependence on back-gate voltage, features of a metallic state[13,48]. The sample resistance shows a sharp drop ~ 1.2 K and vanishes subsequently signaling a transition to the superconducting state with a $T_c$ ~ 920 mK. We confirm the 2D nature of the superconductivity from the anisotropy in magneto-transport with B-field orientations and the presence of the BKT phase transition. Magneto-transport measurements reveal a parallel critical-field manifold of the Pauli limit, possibly due to the presence of high spin-orbit coupling in our system, making it a potential candidate for superconducting circuit elements operational at higher magnetic fields. The sample also exhibits a transition to Bose metallic state a signature of clean 2D superconducting system.

**Results & discussions**

Fig. 1 (a) shows HR-TEM image with the selected area electron diffraction (SAED) pattern of a representative 1T-$MoS_2$ sample prepared from the 2H phase by the plasma treatment[48]. The sample shows extended 1T region, whose area is primarily limited by the area of the starting 2H flake. The sharp SAED spots are indicative of highly crystalline extended 1T regions in our sample. We also find tiny isolated 2H regions (bounded in green) which we believe are phase transformed back as a result of prolonged exposure of energetic electron beam. The 1T phase is characterized by an octahedral coordination between the Mo and S atoms[50]. The top right inset in Fig. 1 (a) shows magnified view of the HR-TEM image of the 1T region while the corresponding atomic arrangements of the Mo and S atoms is shown in the bottom inset. Mo atoms, by virtue of its superior visibility in the TEM images, appear as bright spots in contrast to S atoms[51]. More details on the sample preparation, structural and electrical characterization can be found elsewhere[48].

The optical image of a 1T-$MoS_2$ device on Si/$SiO_2$ substrate is shown in Fig. 1(b). The electrical contacts are defined by photolithography followed by Cr/Au metallization. We estimate a thickness of ~8 nm (~12 layers) for the 1T-$MoS_2$ flake using atomic force microscopy (AFM). A height-profile obtained from the AFM along the red-arrowed line is shown in the inset. Fig. 1 (c) shows the 4P I-V characteristics of the sample at 300 K (red) and 4 K (blue). We use probes 7, 8, 9 and 13 for all the transport measurements presented in this manuscript. The I-V characteristics show excellent linearity both at 300 K (red) and 4 K (blue). The traces are offset along the y-axis by 20 µV for better visibility. We observe that the normal state resistance of the sample at 300 K , ~ 70 Ω, undergoes a slight increase, by ~ 2 Ω , as the sample is cooled down to 4 K. In addition, the resistance of the sample shows little variation with the back-gate voltage $V_g$, as shown in the top-left inset to Fig.1 (c). The linear I-V characteristics void-off any gate-voltage dependence are characteristic suggest the metallic nature of the 1T. From the 4P resistance ~ 70 Ω, we calculate a sheet resistance $R_s$ ~ 108 Ω/☐ for the sample. From the conductivity $\sigma_s = \frac{n_s e^2 \tau}{m^*}$, we obtain $k_F.l = \frac{(h/e^2)}{R_s}$ ~ 239 for our sample which satisfies the Ioffe-Regel criteria[52], $k_F.l \geq 1$, the disorder limit for metallic conduction. $k_F$ is the Fermi wave vector and $l$ is the mean-free-path of electrons in our sample. From the Hall measurements and the sheet resistance we estimate a carrier concentration of ~ $2.3 \times 10^{15}$ cm$^{-2}$ and a mean-free path $l$ ~ 20 nm.

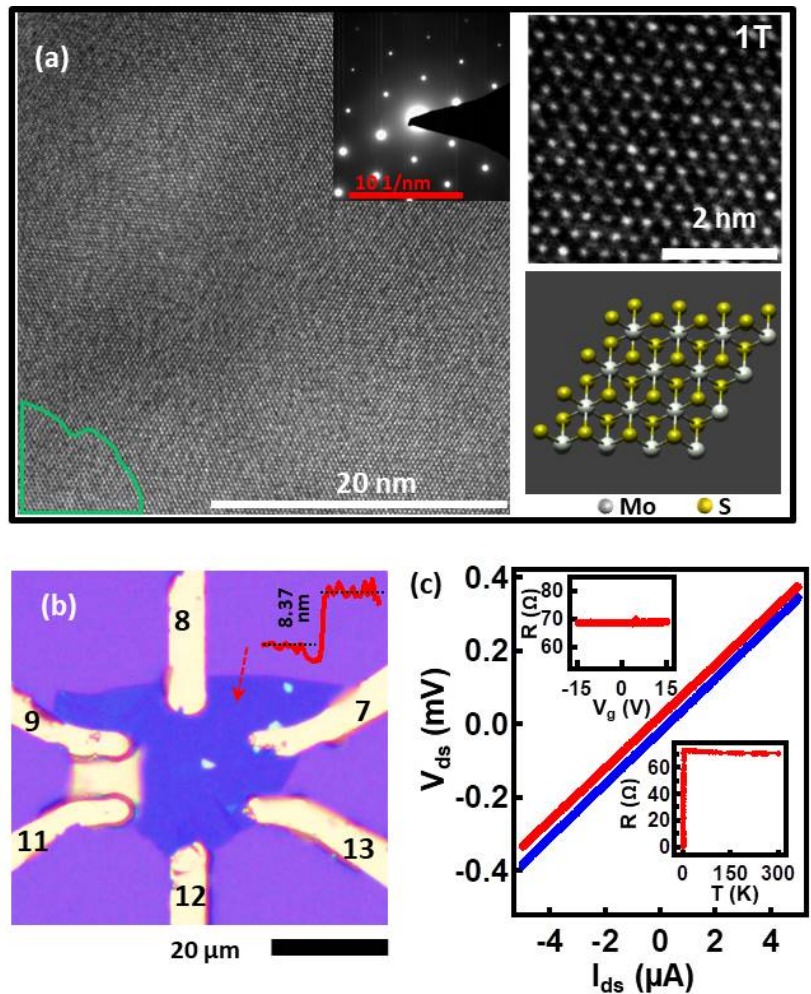

*Figure 1: (a) HR-TEM image with diffraction pattern of a plasma treated $MoS_2$ sample showing the 1T-phase. Area of the 2H-phase is enclosed in green; magnified view of the 1T region and the corresponding atomic arrangement are shown in the right top and bottom insets respectively. (b) The optical image of the device with the probes marked. AFM line profile taken along the arrowed line is given in the top-right showing a thickness of 8.37 nm. (c) 4P I-V characteristics of the device at 300 K (red) and 4 K (blue). The traces are offset along the y-axis by 20 $\mu$V the for visibility. Top-inset: 4P resistance vs back-gate voltage at 300 K, bottom inset: 4P Resistance vs temperature from 300K down to 12 mK.*

The resistance as a function of temperature from 300 K down to 12 mK is shown in the lower-right inset of Fig. 1 (c). As observed from the I-V characteristics the resistance show negligible temperature dependence between 300 K and 4 K negating any hoping-mediated or activated transport observed in disordered systems[13,53]. This also concurs with the fact that the sample consists of continuous 1T regions as inferred from the TEM analysis in contrast to isolated 1T islands; resistance of 1T-MoS$_2$ samples prepared by chemical intercalation routes generally consist of 1T islands in 2H matrix and shows strong temperature dependence in contrast[13]. As the sample is cooled below 4 K the resistance decreases, drops sharply ~ 1.2 K and subsequently vanishes signaling the presence of a superconducting state. We note here that the carrier concentration of our sample is two-orders higher in magnitude compared to the minimum carrier concentration reported for the observation of 2D superconductivity in ionic-liquid gated MoS$_2$[29]. In addition, the sheet resistance of our sample is considerably lower than the resistance quantum, h/(2e)$^2$, the upper limit to observe 2D superconductivity[3].

To understand the nature and dimensionality of superconductivity in our device, we explore magneto-transport in the parallel and perpendicular B-field orientations. Fig. 2 (a) and (b) show the resistance of the sample vs temperature for various B-field values in the parallel and perpendicular orientations respectively. Fig. 2 (c) and (d) shows the magneto-resistance of the sample for temperatures between 12 mK and 1.2 K in the parallel and perpendicular B-field directions respectively. The temperature at which the resistance reduces to 50% of the normal state resistance $R_N$ (~ 70 Ω), at zero B-field, is regarded as the critical temperature $T_c$ ~ 920 mK. $T_c$ for our sample corresponds to a Bardeen-Cooper-Schrieffer (BCS) energy gap $2\Delta_{BCS}$= 3.52 $K_B T_c$ ~ 280 µeV[54]. The transition moves to lower temperature values with B-field and eventually vanishes above a critical field $B_{c2}^{\parallel}$ for parallel and $B_{c2}^{\perp}$ for perpendicular orientations. The critical field is estimated as the field at which the resistance of the sample reduces to 50% of $R_N$. We observe a large anisotropy in the transport with respect to the B-field orientations. In addition, we also observe that for the perpendicular orientation the transition broadens even at very small B-field values compared to that for the parallel configuration. The broadening of the transition may be caused by phase fluctuations induced by the B-field [55]. These observations are atypical of a 3D superconducting state and points towards the presence of a 2D superconducting state in our sample.

For a 2D superconductor, the Tinkham's model predicts the behavior of critical fields for parallel and perpendicular configurations; $B_{c2}^{\parallel}(T) = \frac{\sqrt{3}\phi_0}{\pi\xi_{GL}(0)d_{sc}}\left(1 - \frac{T}{T_c}\right)^{1/2}$ in the parallel orientation and $B_{c2}^{\perp}(T) = \frac{\phi_0}{2\pi\xi_{GL}(0)^2}\left(1 - \frac{T}{T_c}\right)$ in perpendicular orientation[28], where $\phi_0$ is the flux quantum, $\xi_{GL}(0)$ is the Ginzburg-Landau coherence length and $d_{sc}$ is the effective thickness of the superconductor. Fig. 2 (e) shows the dependence of the critical B-field on the temperature for parallel (green) and perpendicular (blue) configurations. The green and blue dotted lines are fits using Tinkham's model for the parallel and perpendicular orientations respectively, showing good agreement with the theory. From the fits, we estimate $\xi_{GL}(0) \sim$ 11.9 nm and $d_{sc} \sim$ 12.1 nm. The BCS coherence length, $1.35\,\xi_{GL}(0) < l$ implies that our device can be regarded as a clean superconductor. The estimated superconducting thickness obtained from the fit is of the same order but slightly larger than the sample thickness measured by AFM; similar discrepancies have been reported in the past[22,28]. We note here that in the parallel field configuration the critical field is very sensitive to small angular deviations in the sample orientation which may also contribute to this error[54].

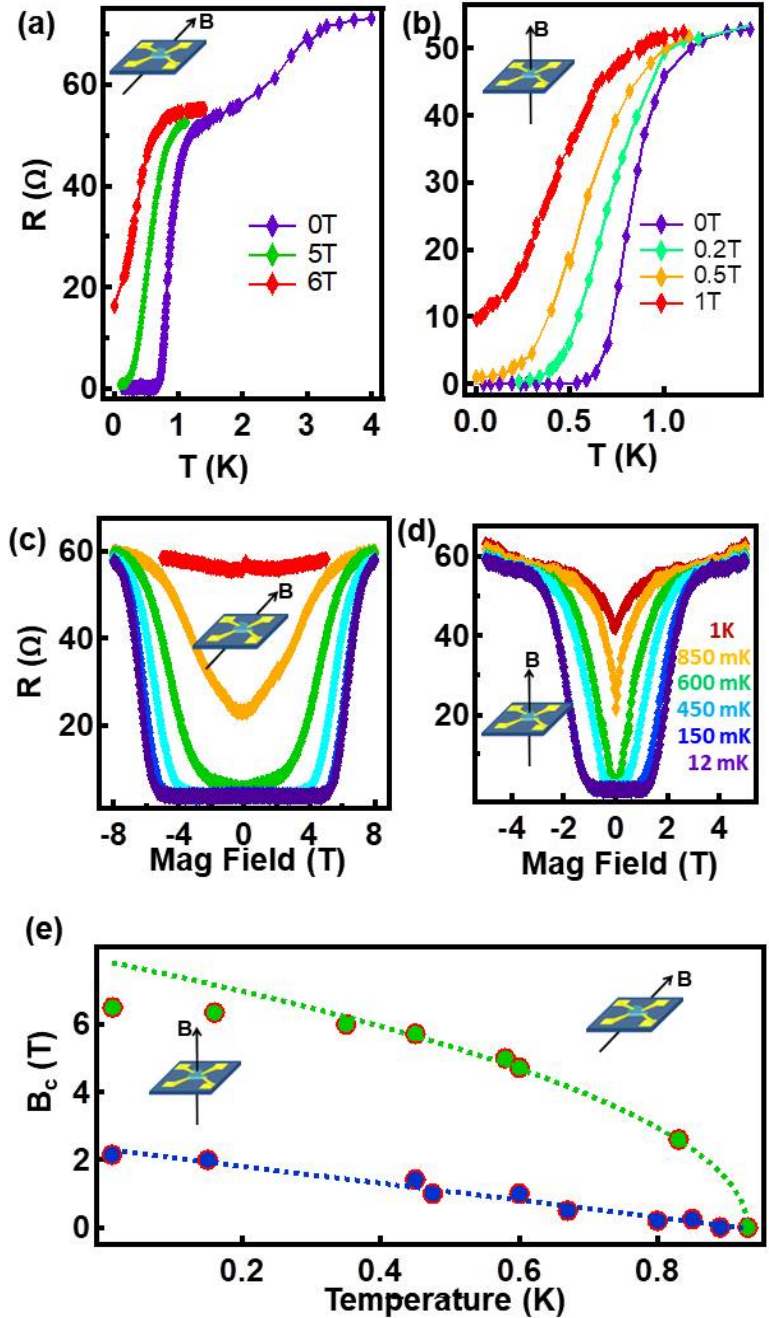

*Figure 2: (a) Sample resistance vs temperature for parallel B-fields in the superconducting transition region. (b) The resistance versus temperature at different B-fields in perpendicular orientation. (c) & (d) The Magneto-resistance at different temperatures between 12 mK and 1 K with B-field in parallel (c) and (d) perpendicular orientations. (e) Critical field verses temperature for parallel (green) and perpendicular (blue) orientations. The dotted lines show fit to the Tinkham's model.*

Pauli limit[54] described by the BCS theory put an upper bound to $B_{c2}^{\parallel}(0) = 1.86\,T_c$ in 2D superconductors. In our sample, the measured $B_{c2}^{\parallel}$ at 12 mK is $\sim$ 6.25 T, ~four-times the value 1.618 T predicted by the Pauli limit. A similar enhancement in the $B_{c2}^{\parallel}$ has been observed in systems with high SOC (i) TMDC superconductors in the monolayer limit owing to Ising pairing of electrons[29-31] (ii) crystalline thin-film superconductors with high SOC and substrate induced symmetry breaking[25] and, (iii) recently in centro-symmetrtric few layer TMDCs[32]. This has been suggested due to the SOC induced spin splitting of conduction electrons making pair breaking unfavorable under external magnetic field; further studies are required to understand the detailed mechanism.

A superconducting state is characterized by a critical current $I_c$, above which the material turns normal [54]. Fig. 3 (a) shows the I-V characteristics from 12 mK (blue) to 4 K (red) at zero B-field. A plot of $I_c$ vs temperature is shown in the inset. We observe that as the $I_c$ reduces with temperature and finally becomes zero as the normal state is reached. Dynamics of vortices leave clear signature in the transport characteristics. In a 2D superconductor at zero magnetic field, the transition to zero resistance involves (i) formation of Cooper-pairs at $T_c$ and, (ii) vortex-antivortex pairing reducing dissipation, the BKT transition and, (iii) the resistance drops to zero for subsequently lower temperatures following the condensation of the Cooper-pairs to single state. BKT phase transition is a hallmark phenomenon of 2D superconductivity. A signature of the BKT transition is the deviation from the linear nature of the I-V in the normal state to a power-law dependence $V \propto I^{\alpha(T)}$ obeying Halperin-Nelson scaling as a result of the motion of the free vortices in the system[56]; the exponent $\alpha(T)$ takes a critical value of 3 at the transition[22,35,56]. Fig. 3 (b) shows the log-log scale plot of the I-V characteristics in the vicinity of the superconducting transition exhibiting the power law dependence. The top inset shows a plot of the scaling exponent $\alpha(T)$ vs temperature as the sample undergoes a transition from the normal to the superconducting state. As the temperature is lowered the slope $\alpha$ increases from unity corresponding to the Ohmic behavior and diverges rapidly around α = 3. The grey dotted line in the main panel and the purple dotted-line in the inset correspond to α = 3 and, the corresponding temperature is the BKT transition temperature $T_{BKT}$ ~ 675 mK.

We observe a hysteresis in the I-V traces between forward and reverse current sweeping directions in the superconducting state as shown in the upper panel in Fig. 3(c). This is regarded as another signature of the presence of vortex-antivortex pairs in the system[22,57]. Above $T_{BKT}$ the free vortices are stable and the correlations between distant pairs are destroyed. Below $T_{BKT}$ the free vortices created as a result of the driving current also recombine to pair, and the voltage developed due to the vortex motion in the unbound state causes the hysteresis in the critical current $\Delta I_c$ [57,58].

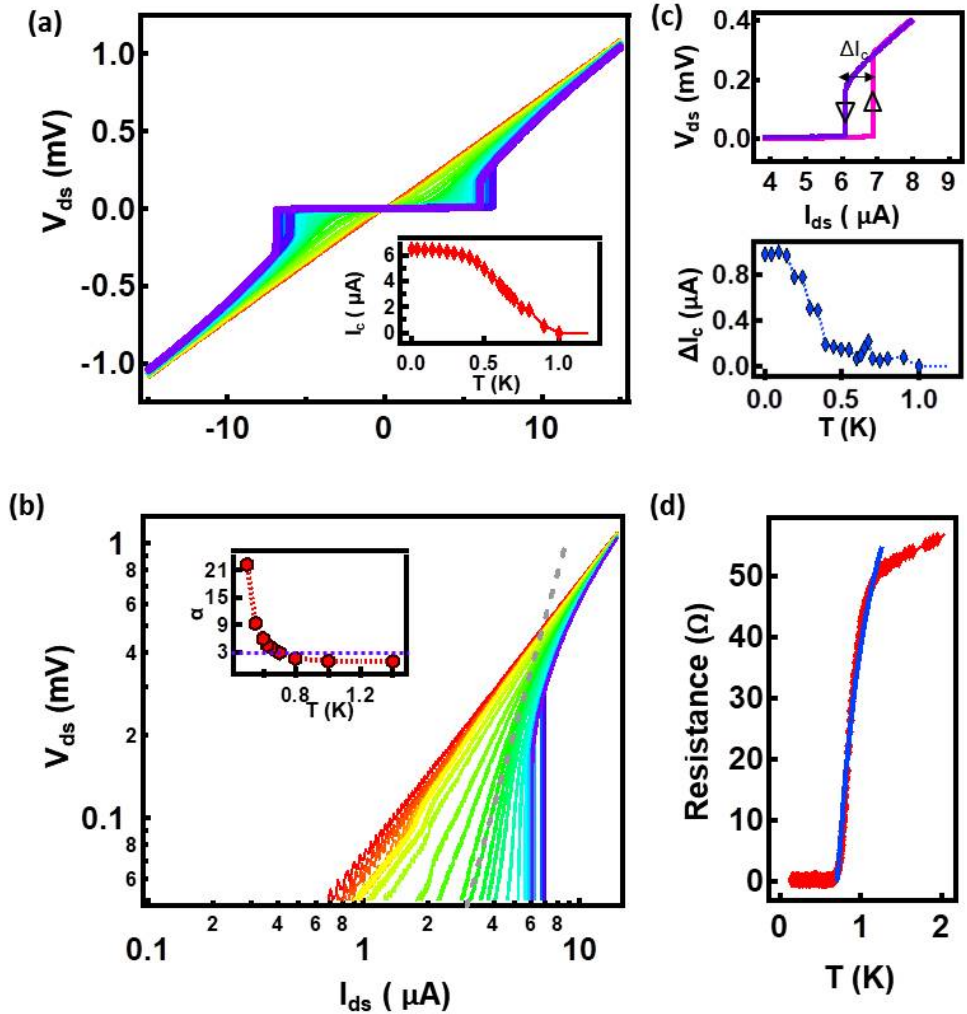

*Figure 3: (a) The I-V characteristics for temperatures 12 mK (blue) to 4 K (red) at 0 T. Inset: The plot of critical current versus temperature. (b) The I-V plot in the log scale showing the power law behaviour $V \propto I^{\alpha(T)}$, where the grey dotted line corresponds to α = 3 at a temperature ~ 0.675 K corresponding to the BKT transition. Inset shows the value of the scaling exponent, $\alpha(T)$ versus temperature. (c) Top: magnified view of the I-V characteristics at 12 mK showing hysteresis with sweep direction Bottom: Plot of hysteresis in $I_c$ verses temperature (d) Plot of Resistance versus temperature at 0 T where the blue line shows fit to the Halperin-Nelson equation, giving an estimate of $T_{BKT}$ = 0.689 K.*

$\Delta I_c$ as a function of temperature is shown in the lower panel of Fig. 3 (c). The hysteresis reduces with temperature and disappears ~ 675 mK which corresponds to the $T_{BKT}$ estimated from the I-V characteristics using the scaling theory.

According to Halperin-Nelson equation[59,60] the resistance in a 2D superconductor in the vicinity of the transition varies as $\exp(-b(T - T_{BKT})^{-\frac{1}{2}})$. A plot of resistance vs temperature with the fit to the equation (blue line) is shown in Fig. 3 (d). We extract a value of ~ 689 mK for the $T_{BKT}$ from the fit which is consistent with the that obtained from the I-V characteristics.

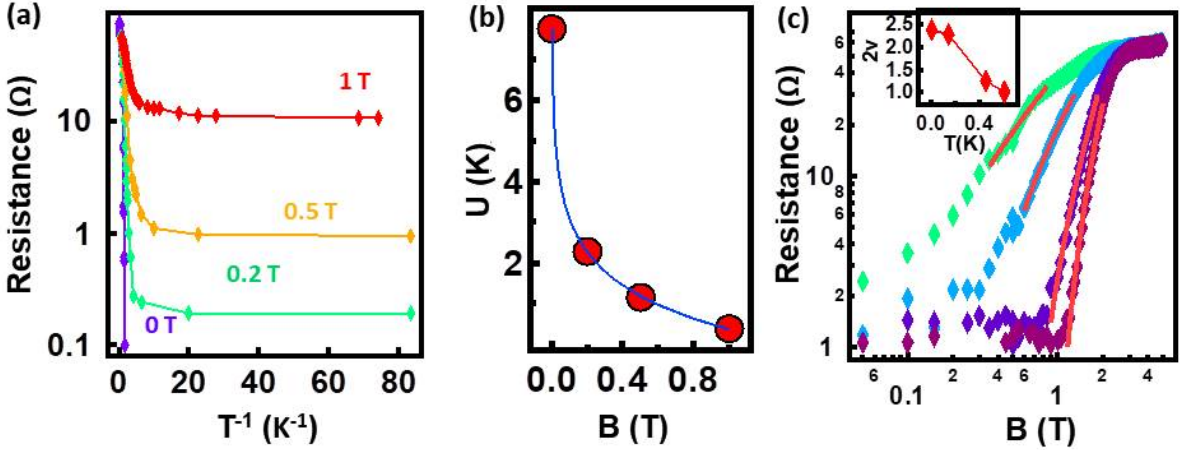

*Figure 4: (a) Arrhenius plot of resistance versus temperature for various perpendicular B-fields. (b) The red dots are the vortex binding energy calculated from linear region in (a) and the blue line shows the fit to equation $U(B) = U_0 ln (B_o/B)$. (c) The magneto-resistance plot for the perpendicular B-field in log-scale. The solid red lines on the traces are fit to the scaling relation $R \propto (B - B_{c0})^{2\nu}$, a plot of $2\nu$ verses temperature is given in the inset.*

A strongly disordered 2D superconductor makes a transition to an insulating state under the application of a perpendicular B-field[24]. In contrast, systems with low disorder undergoes a quantum phase transition to an intermediate 2D metallic state with a resistance much lower than $R_N$[22,24]. This state is characterized by a saturation of the resistance at temperatures far below $T_c$. Akin to thermally driven BKT transition, the guage fluctuations caused by the B-field destroys phase-coherence and the uncondensed Cooper-pairs and free vortices cause dissipation in the system. An Arrhenius plot of resistance for perpendicular B-fields below $B_{c2}^{\perp}$ is shown in Fig. 4(a). Though a sharp drop in the resistance is observed at the superconducting transition point, the system does not exhibit a zero resistance state for finite field values as the temperature is lowered and saturates to small values compared to $R_N$. The resistance exhibit an activated behavior for temperatures close to $T_c$ due to temperature driven depairing of vortex-antivortex pairs. The pair dissociation energy $U(B)$ is obtained from the fit to Fig. 4(a) using $R \propto e^{-\frac{U(B)}{T}}$ [22]. Fig. 4(b) shows a plot of U(B) hence obtained vs B-field with a fit to the equation $U(B) = U_0 ln (B_0/B)$ [22], shown in blue. From the fit we obtain $U_0$ = 1.18 K ~ 1.76 $T_{BKT}$ and $B_0$ = 1.359 T ~ $B_{c2}^{\perp}$.

The superconducting to the Bose metallic state transition is characterized by a resistance scaling $R \propto (B - B_{c0})^{2\nu}$ where $B_c (0)$ is the critical field for the superconductor to Bose metal transition and ν is the exponent by which the superfluid correlation length diverges[22,24]. Fig. 4(c) is a log-log scale plot of resistance vs perpendicular B-field for temperatures below $T_{BKT}$ with a fit to the above equation. A plot of the scaling exponent 2ν extracted from the fit vs temperature is shown in the inset. The exponent assumes a value of unity near $T_{BKT}$ corresponding to a linear behavior and grows nonlinear as temperature is reduced with 2ν ~ 2.3 for lower temperatures. We note that the exponent has been shown to vary with sample thickness and the value of the exponent we obtained is comparable to those observed elsewhere for similar systems[22].

**Conclusions**

In this manuscript, we explored 2D superconductivity in a large area few-layer 1T-MoS$_2$ sample. We investigated transport properties of the sample from 300 K down to 4 K in a closed cycle cryostat and from 4 K down to 12 mK in a dilution refrigerator. The carrier concentration of ~ 2.3 X 10$^{15}$ cm$^{-2}$ and a sheet resistivity of ~ 108 Ω/☐ put the sample in the clean metallic regime. In addition, the linear I-V characteristics with no dependence on the back-gate voltage also underline metallic conduction in our sample. Below a temperature of ~ 1.2 K the sample undergoes a superconducting transition with a $T_c$ ~ 920 mK. We observe a clear anisotropy in the magneto-transport with respect to field orientations obeying Tinkham's model. The superconductivity in our sample is accompanied by BKT phase transition, enhanced parallel critical-field and a transition to a Bose metal state, all of which are landmark features of clean 2D superconducting system. The BKT transition is identified from the exponential scaling of the I-V characteristics, hysteresis in the critical current and also from resistance vs temperature characteristics obeying Halperin-Nelson equation. The sample undergoes a transition from the superconducting to a 2D quantum metallic state, the Bose metal, under the application perpendicular B-field which is generally observed in systems with low disorder density. We observed a manifold increase in the parallel upper critical field beyond the Pauli limit possibly due to high SOC in our sample.

The metallic polymorphic phase of MoS$_2$ is a widely studied system from the materials point of view, though, in-depth electrical characterization are in wanting. We believe this work will add depth to the current understanding of the nature of transport in the system. The enhanced upper critical-field, even in the multilayer regime, makes 1T-MoS$_2$ a potential system for on-chip circuit elements and cavity QED applications requiring operation at higher external magnetic fields. The possibility of scalably and controllably engineering the 2D superconducting 1T phase on the semiconducting 2H phase opens up a new landscape for hybrid quantum device circuits.

**Materials and Methods**

MoS$_2$ samples commercially procured from SPI supplies are exfoliated and transferred onto a clean Si wafer hosting 300 nm SiO$_2$ layer. The 2H to 1T phase conversion is achieved by exposing the exfoliated samples to Forming gas (90% Ar + 10 % H2) microwave plasma. The details of the plasma system and the 1T conversion process can be found elsewhere[48]. Contacts on the 1T samples are made by standard photo lithography followed by Cr/Au metallization. Low noise transport measurements are performed in a cryogen-free dilution refrigerator equipped with a superconducting magnet and also in a closed cycle 4 K cryo-cooler.

**Acknowledgements:** The authors acknowledge IISER Thiruvananthapuram for the infrastructure and experimental facilities and Anil Shaji for in-depth discussions and a critical reading of the manuscript. M.T. acknowledges the funding received from DST-SERB extramural program (SB/S2/CMP-008/2014). C.H.S. acknowledges CSIR and A.P.S. acknowledges INSPIRE for the fellowship.

**Author contribution:** MT conceived the problem. APS prepared the samples and performed the structural characterizations, CHS and APS fabricated device, CHS performed the transport measurements and analyzed the data, CHS, SSV designed the data acquisition system and, CHS and MT co-wrote the manuscript.